\documentclass[prb,aps,twocolumn,showpacs]{revtex4}
\usepackage{amsmath,amssymb,amsfonts,float}
\usepackage[dvips]{graphicx,color}    
\usepackage{times}       
\usepackage{pifont}
 
\newcommand{\bolk}{\mathbf{k}}
\newcommand{\bolq}{\mathbf{q}}
\newcommand{\bole}{\mathbf{e}}
\newcommand{\bolQ}{\mathbf{Q}}

\newcommand{\bolr}{\mathbf{r}}

\newcommand{\ket}[1]{| #1 \rangle}  
\newcommand{\VEV}[1]{\langle #1 \rangle}  

\newcommand{\be}{\text{e}}

\newsavebox{\dotdot}
\savebox{\dotdot}[3mm]{\shortstack{\circle*{0.8}\\ \\ \circle*{0.8}}}

\begin{document}
\title{
Magnon BEC and various phases of 3D quantum helimagnets  
under high magnetic field
}
\author{Hiroaki~T.~Ueda$^{1,2}$ and Keisuke~Totsuka$^1$}
\affiliation{$^1$ Yukawa Institute for Theoretical Physics, 
Kyoto University, Kitashirakawa Oiwake-Cho, Kyoto 606-8502, Japan\\
$^2$ Condensed Matter Theory Laboratory, RIKEN, Wako, Saitama 351-0198, Japan}
\begin{abstract}
We study high-field phase diagram and low-energy excitations of 
three-dimensional quantum helimagnets.   
Slightly below the saturation field, 
the emergence of magnetic order may be viewed as Bose-Einstein condensation (BEC) 
of magnons. The method of dilute Bose gas enables a quantitative 
analysis of quantum effects in these helimagnets and thereby three 
phases are found: cone, coplanar fan and a phase-separated one. 
As an application, we map out the phase diagram of a 3D helimagnet 
which consists of frustrated $J_{1}$-$J_{2}$ chains 
as a function of frustration and an interchain coupling. 
Moreover, we also calculate the stability of the 2-magnon bound state 
to investigate the possibility of the bound-magnon BEC.
\end{abstract}            
\maketitle
\section{introduction}
Magnetic frustration introduces several competing states which are 
energetically close to each other and thereby destabilizes 
simple ordered states.  
One way to compromise two or more competing orders is to 
assume a helical (spiral) spin structure\cite{Yoshimori}. 
In this paper, we discuss the high-field behavior of 
a spin-1/2 Heisenberg model with generic interactions: 
\begin{equation}
H=\sum_{\langle i,j\rangle}J_{ij}\,{\bf S}_i{\cdot}{\bf S}_j 
+ \text{H}\sum_{j}S^{z}_{j} .
\end{equation}
For the simplest case with one magnetic ion per unit cell, one can 
easily find the classical ground state by minimizing the Fourier 
transform of the exchange interactions:
\begin{equation}
\epsilon(\bolq)=\sum_{j} \frac{1}{2} J_{ij}\cos \left(
\bolq {\cdot}(\bolr_i-\bolr_j)\right)\ ,
\end{equation}
where the summation is taken over all $j$-sites connected to 
the $i$-site by $J_{ij}$.    
When $\epsilon(\bolq)$ takes its minima $\epsilon_{\text{min}}$ 
at $\bolq=\pm \bolQ$, helical order with the wave number $\bolQ$ 
or $-\bolQ$ appears 
($\pm \text{\bf Q}$ are not equivalent to each other).  

\begin{figure}[H]
\begin{center}
\includegraphics[scale=0.4]{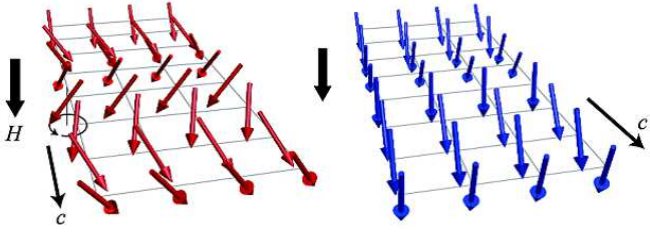}
\caption{(Color online) Two spin structures considered here: 
`cone' (left) and `fan' (right). 
In the fan structure, the spins are lying in a single plane 
({\em coplanar}). The ${\bf Q}$-vector is pointing along 
the $c$-axis. 
\label{fig:cone-fan}}
\end{center}
\end{figure}

When the external magnetic field is perpendicular 
to the spiral plane, the spiral is smoothly deformed 
into the so-called {\em cone} state (Fig.\ref{fig:cone-fan}) 
and this persists until all spins eventually get polarized 
at the saturation field $\text{H}_{\text{c}}$.  
When the system has an easy-plane anisotropy and 
the external field is applied in the 
spiral plane, on the other hand, the system undergoes a (first-order) 
metamagnetic transition into a coplanar {\em fan} 
phase\cite{Nagamiya}. 

One of the simplest models which exhibit, at least in the classical 
limit, the helical order is a three-dimensionally coupled 
Heisenberg chains with nearest-neighbor- 
(NN) $J_1$ and next-nearest-neighbor (NNN) $J_2$ coupling. 
Because of strong quantum fluctuation in one dimension, 
the spin-singlet ground state of a single decoupled $S=1/2$ chain can be 
quite different\cite{Tonegawa,higher-S-J1J2} from its classical counterpart.  
However, it is generally expected that interchain couplings 
may eventually stabilize the classical helical order. 
In fact, many compounds which contain these 1D-chains 
as subsystems and display magnetic long-range orders are known 
(see, for instance, TABLE I. in Ref.\onlinecite{hase}).  
Despite this naive expectation, even relatively mild quantum 
fluctuations in three dimensions may destabilizes the classical 
ground state in some frustrated systems\cite{Nikuni-Shiba-1,Nikuni-Shiba-2}.  
Therefore, it would be interesting to explore the possibility 
that quantum fluctuation replaces the classical cone state with 
other stable ones e.g. a coplanar fan.  

Another interesting feature peculiar to the quantum case is that 
for a region slightly below the saturation field, 
we can view the emergence of various kinds of (weak) magnetic order 
as Bose-Einstein condensation (BEC) of magnons which enables us 
to use the full machinery of many-body theories\cite{HuaShi}.  
The concept of magnon BEC has been successfully applied to 
explain various experimental 
results\cite{Giamarchi-R-T-08,Nikuni-O-O-T-00,CsCuCl4-BEC}.  
By using dilute-Bose-gas approach, Batyev and Braginskii\cite{Batyev-1} 
discussed magnetic structure near saturation  
from a general point of view and concluded that  
this is the case if a certain condition for 
the bosonic interactions is satisfied. 

Recently, helimagnetism attracts renewed interest
in the context of  
multiferroicity\cite{MF-review} and multiferroic behavior 
has been reported for various helimagnets.  
For example, a helimagnetic material LiCuVO$_4$ may be viewed as 
coupled quantum $S=1/2$ $J_{1}$-$J_{2}$ chains and, as is expected 
from the classical theories, 
exhibits helical spin order\cite{LiCuVo4} 
and ferroelectricity\cite{LiCuVO4-MF-Naito,LiCuVO4-MF-Schrettle} 
simultaneously 
under moderate magnetic field. 
When the field is very high, on the other hand, this compound shows 
modulated collinear order\cite{LiCuVo4-mag}, 
which contradicts with 
the aforementioned classical prediction\cite{Nagamiya}, 
and this suggests that quantum fluctuation plays an important role.  
In these multiferroic materials, external magnetic field 
provides us with a way of controlling 
polarization\cite{LiCuVO4-MF-Schrettle} and it is crucial 
to understand magnetic structures in high magnetic 
field.  However, except for one-dimensional cases, only few 
reliable results are known for quantum systems so far.  
Our aim in this paper is to determine the stable spin configurations 
of 3D spin-1/2 helimagnets in a fully quantum-mechanical manner.  

The organization of the present paper is as follows. 
In Sec.~\ref{Sec:General}, 
we describe how magnon BEC technique is used to determine possible 
magnetic structures slightly below the saturation field. 
Our dilute Bose gas approach predicts that in general 
there are at least three types of quantum phases 
({\em cone}, {\em fan} and an attraction dominant phase; 
see FIG.\ref{fig:cone-fan}) in the high-field region. 
By mapping to an effective Lagrangian, we study the low-energy 
properties of the cone and the fan phases in Sec.~\ref{Sec:low-energy} 
and show that, on top of the standard Goldstone mode, there is 
a yet another gapless mode in the fan phase which corresponds to 
translation. 

In general, when ${\bf Q}$ is commensurate with the underlying 
lattice, lattice symmetry allows several higher-order interactions 
which may pin the above gapless translational motion.  
In Sec.~\ref{Sec:Commensurability}, we discuss the effects of 
commensurability on these phases. 

As a concrete example, the phase diagram of 
a model of coupled $S=1/2$ $J_1$-$J_2$ chains 
($J_1$-$J_2$-$J_3$ model) is considered in Sec.~\ref{Sec:J1J2model}. 
In this model, we also study the stability of 2-magnon bound state. 
If the bound-magnon BEC occurs, the transverse magnetic moment vanishes. 
As a result, we find that, on top of the above three phases, a spin nematic phase appears.

\section{General Formalism}
\label{Sec:General}
\subsection{Mapping to dilute-Bose gas}
To apply the powerful Bose-gas technique, we first rewrite the original 
spin model in terms of bosons. 
When the external field H is larger than the saturation field 
H$_{\text{c}}$, spins are fully polarized along H (downward, here) 
and any spin-flip excitations from this reference state can be expressed 
exactly in terms of hardcore bosons.  
Specifically, we write the spin operators as: 
\begin{equation}
S^z_l=-1/2+\beta^\dagger_l\beta_l \; , \;\;
S_l^+ =\beta_l^\dagger \; , \; \;  S_l^- =\beta_l \; .
\end{equation}
Then, the original spin Hamiltonian in general may be rewritten 
in the following way:
\begin{equation}
H = \sum_{q}(\omega (\bolq) - \mu)\beta^\dagger_{\bolq}\beta_{\bolq}
+\frac{1}{2N}  \sum_{\bolq,\bolk,\bolk^\prime}  V_{\bolq} 
\beta_{\bolk+\bolq}^\dagger\beta_{\bolk^\prime-\bolq}^\dagger
\beta_{\bolk}\beta_{\bolk^\prime},\label{Hboson}
\end{equation}
\begin{equation}
\begin{split}
&\omega(\bolq) =\epsilon(\bolq)-\epsilon_{\text{min}}\ ,\ \ \ 
\mu={\rm H}_{\text{c}}-{\rm H}\ ,\\
&{\rm H}_{\text{c}}=\epsilon({\bf 0})-\epsilon_{\text{min}}\ ,\ \ \ 
V_{\bolq} =2(\epsilon(\bolq)+U)\ ,
\end{split}
\end{equation}
where $N$ is the number of lattice sites. 
In helimagnets, the single-spin flip excitation $\epsilon({\bf q})$ 
takes its minima $\epsilon_{\text{min}}$ at two inequivalent 
${\bf q}$-points $\pm{\bf Q}$.  
The external field H controls the chemical potential $\mu$ 
and the on-site interaction $U(\rightarrow \infty)$ has been added 
to impose the hardcore constraint.  

In what follows, we consider a cubic lattice and 
assume that helical- and ferromagnetic/antiferromagnetic 
order occur along the $c$-axis 
and in the $ab$ plane, respectively 
(i.e. $\mathbf{Q}=(0,0,Q)$ or $(\pi,\pi,Q)$).  
Also, in order to avoid confusion, we use 
the indices $(a,b,c)$ for the real-space coordinate and 
reserve $(x,y,z)$ for the spin directions.  

We see that magnon BEC occurs 
when the external field is smaller than the saturation 
field: $\text{H} < \text{H}_{\text{c}}$ ($\mu>0$).  
Although the hard-core formulation is valid only for spin-1/2, 
it can be generalized\cite{Batyev-2}, with a little modification, 
to arbitrary spin-$S$.
\subsection{Ginzburg-Landau analysis}
\begin{figure}[H]
\begin{center}
\includegraphics[scale=0.5]{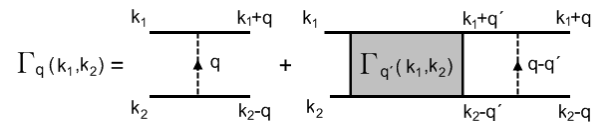}
\caption{Ladder approximation to the interaction vertex.%
\label{Fig:ladder}}
\end{center}
\end{figure}

The thermal potential per site $E/N$ of the dilute Bose gas 
is determined by the interaction among 
the condensed bosons at ${\bf q}=\pm {\bf Q}$ 
and the ground-state Boson densities $\rho_{\pm {\bf Q}}$ are obtained 
by minimizing $E/N$. 
If we denote the renormalized interactions between 
the same bosons and that between different ones respectively 
as $\Gamma_1$ and $\Gamma_2$, the energy density $E/N$ is given by
\begin{equation}
\begin{split}
\frac{E}{N} &=\frac{1}{2}\Gamma_1
\left(\rho_{\bolQ}^2+\rho_{-\bolQ}^2\right)
+\Gamma_2 \, \rho_{\bolQ} \rho_{-\bolQ} 
- \mu(\rho_{\bolQ}+\rho_{-\bolQ}), \\
&= \frac{1}{4}(\Gamma_1+\Gamma_2)(\rho_{\bolQ} +\rho_{-\bolQ})^{2} 
+ \frac{1}{4}(\Gamma_1-\Gamma_2)(\rho_{\bolQ}-\rho_{-\bolQ})^{2} \\
& \phantom{=} -\mu (\rho_{\bolQ} +\rho_{-\bolQ})
\end{split}
\label{EffPotential}
\end{equation}
where $\rho_{\bolq}=|\VEV{\beta_{\bolq}}|^2/N$.
First we note that the energy function $E/N$ has discrete 
$\mathbb{Z}_{2}$-symmetry $\mathbf{Q}\leftrightarrow -\mathbf{Q}$. 
When the external field H is sufficiently close to 
the saturation field H$_{\text{c}}$, we may expect that 
the condensed boson is dilute 
and one can safely use 
the ladder approximation\cite{HuaShi,footnote1}  
to calculate the interaction vertex (see Fig.\ref{Fig:ladder}):  
\begin{multline}
\Gamma_{\bolq} (\bolk_1,\bolk_2)  \\
=\!V_{\bolq}
- \frac{1}{N}\sum_{\bolq^\prime}\!\frac{\Gamma_{\bolq^\prime}(\bolk_1,\bolk_2)
V_{\bolq-\bolq^\prime}}{\omega(\bolk_1{+}\bolq^\prime)
+\omega(\bolk_2{-}\bolq^\prime)-\omega(\bolk_1)-\omega(\bolk_2)}.
\label{laddereq}
\end{multline}
From this, one obtains the parameters $\Gamma_{1}$ and $\Gamma_{2}$  
as\cite{Nikuni-Shiba-2} 
$\Gamma_1 =\Gamma_0(\bolQ,\bolQ)\ ,\ 
\Gamma_2 =\Gamma_0(\bolQ,-\bolQ)+\Gamma_{-2\bolQ}(\bolQ,-\bolQ)
$.
\subsubsection{Cone phase}
Different phases appear according to the values of $\Gamma_{1,2}$. 
If $\Gamma_2 > \Gamma_1 >0$,  
the ground state is given by
$\rho_{\bolQ} =\rho=\mu/\Gamma_1$,  
$\rho_{-\bolQ}=0$ (or vice versa) and 
$E/N=- \mu^2/(2\Gamma_1)$.
Hence, the spin configuration is determined as:
\begin{equation}
\begin{split}
& \VEV{\beta_l}=\sqrt{\rho}\exp\{\pm i({\bf Q}{\cdot}{\bf R}_l \!+\! \theta)\}
\ ,\VEV{S_l^z}=-\frac{1}{2}+\rho\ ,\\
& \VEV{S_l^x} \!=\! \sqrt{\rho}\cos ({\bf Q}{\cdot}{\bf R}_l \!+\! \theta),\ 
\VEV{S_l^y} \!=\! \mp\sqrt{\rho}\sin ({\bf Q}{\cdot}{\bf R}_l \!+\! \theta).
\end{split}
\end{equation}
That is, the cone state (the left panel of Fig.\ref{fig:cone-fan}), 
which exists already in the classical case\cite{Nagamiya},  
is favored for $\Gamma_{2}>\Gamma_{1}$. 

It is easy to see that this phase exhibits the multiferroic behavior.  
According to the so-called spin-current mechanism, 
a microscopic electric polarization 
${\bf P}_{ij}$, which is associated with a pair of sites $i$ and $j$, 
is given by\cite{MF-review,KNB}
\begin{equation}
{\bf P}_{ij} = \eta \, {\bf e}_{ij}\times(\VEV{{\bf S}_i}
\times\VEV{{\bf S}_j})\label{MF-P}\ ,
\end{equation}
where $\eta$ is a constant. 
When the external field H is parallel to the $a$(or $b$)-axis, 
the $\mathbf{Q}$-vector is in the spiral plane, which is perpendicular 
to $\mathbf{H}$, and the summation of the local polarization ${\bf P}_{ij}$ 
over the lattice yields a finite polarization $\propto \rho\, Q\sin Q$ 
parallel to $\mathbf{Q}{\times}\mathbf{H}$. 
When $\mathbf{H}$ is along the $c$-axis, on the other hand, 
the local polarization sums up 
to zero and the system shows no ferroelectricity. 
\subsubsection{Fan phase}
If $\Gamma_1>\Gamma_2$ and $\Gamma\equiv \Gamma_1+\Gamma_2>0$, 
on the other hand, 
the two modes condense simultaneously and 
the ground state is determined as:
$\rho_{\bolQ} =\rho_{-\bolQ}=\rho^\prime=\mu/\Gamma$, 
$\frac{E}{N}=- \mu^2/\Gamma$ 
\begin{equation}
\begin{split}
\VEV{\beta_l}& =\sqrt{\rho^\prime}
\left\{ 
\be^{i({\bf Q}{\cdot}{\bf R}_l+\theta_1)}
+\be^{i(-{\bf Q}{\cdot}{\bf R}_l+\theta_2)}
\right\}  \ ,\\
\VEV{S_l^z}
& =-\frac{1}{2}+4\rho^\prime\cos^2 ({\bf Q}{\cdot} {\bf R}_l
+\frac{\theta_1-\theta_2}{2})\ ,\\
\VEV{S_{l}^{\pm}}
& =2\sqrt{\rho^\prime}\cos ({\bf Q}{\cdot} {\bf R}_l
+\frac{\theta_1-\theta_2}{2})\be^{\mp i\frac{\theta_1+\theta_2}{2}}\ .
\label{superfan}
\end{split}
\end{equation}
The two parameters $\theta_{1}$ and $\theta_{2}$ characterize arbitrary phases 
of the two condensates $\VEV{\beta_{\bf Q}}$ and $\VEV{\beta_{-{\bf Q}}}$, 
respectively and lead to two different low-energy excitations. 
Since ${\VEV{S_l^y}}/{\VEV{S_l^x}}=-\tan\frac{\theta_1+\theta_2}{2}$, 
the spins assume a coplanar configuration ({\em fan}) shown 
in the right panel of FIG.\ref{fig:cone-fan}.  

The ferroelectric property of this phase can be seen again 
from eq.(\ref{MF-P}). 
If one moves from one site to the next along the $c$-axis, 
spins change their direction periodically within a basal plane 
specified by the azimuthal angle $(\theta_1+\theta_2)/2$ 
(see FIG.\ref{fig:cone-fan}).   
Although the vector chirality on each bond 
$(\VEV{{\bf S}_i}{\times}\VEV{{\bf S}_{i+{\bf e}_c}})$ is always 
pointing a fixed direction perpendicular to the basal plane, 
it changes the sign within a period; for the first half period, 
it is positive and for the latter negative.  Hence the local 
polarizations 
$\mathbf{e}_{\text{c}}{\times}(\VEV{{\bf S}_i}{\times}
\VEV{{\bf S}_{i+{\bf e}_c}})$, 
when summed up along the $c$-axis, exactly cancel out and 
yield zero macroscopic polarization 
(note that only bonds parallel to the $c$-axis give non-zero contribution). 

Here we would like to stress that the fan state here does not 
require any kind of easy-plane 
anisotropy and should be distinguished from its classical counterpart  
which exists {\em only} in easy-plane helimagnets\cite{Nagamiya}.   
It is interesting to observe that in the second case ({\em fan}) 
the ordinary superfluid order ($\VEV{S^{\pm}}\neq 0$) and 
the spin-density wave, where $S^{z}_{l}$ modulates with momentum 
$2\bolQ$, coexist. 

\subsubsection{Attraction-dominant phase}
\label{attraction-dominant}
When $\Gamma_1<0$ or $\Gamma_1+\Gamma_2<0$, 
low-energy bosons around $\mathbf{q}=\pm \mathbf{Q}$ 
attract each other. 
If the energy (\ref{EffPotential}) is taken literally, 
first order transitions may be expected on general grounds. 
In some cases, this scenario may be the case 
and, on physical ground, we may expect bosons to "collapse" in real space. 
It might well be that as a subsequent phase a cone or fan phase appears via first-order transition. 
However, eq.(\ref{EffPotential}) is based on the assumption 
that magnon BEC occurs in the single-particle channel and 
may not work when we expect magnon bound states stabilized by 
strong attraction.  
In fact, this conditions for $\Gamma_{1,2}$ implies nothing but 
instability in the one-magnon condensates. 
In Sec.\ref{Sec:J1J2model}, we calculate the energy of the 2-magnon bound state in the concrete model. As a result, we see that the bound state tends to be favored in attraction-dominant phase. 

\begin{figure}[H]
\begin{center}
\includegraphics[scale=0.5]{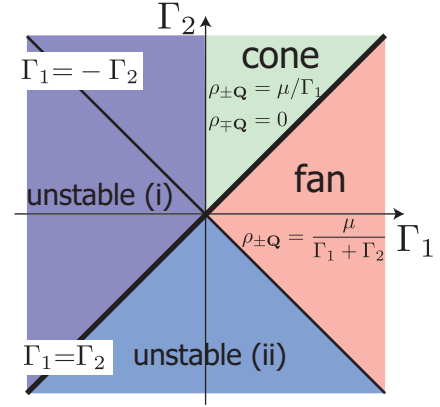}
\caption{(Color online) Phase diagram in $(\Gamma_1,\Gamma_2)$-plane. 
When $\Gamma_{1}<0$ or $\Gamma_1+\Gamma_2<0$, the energy function 
$E/N$ is, at least within single-particle BEC, unstable 
({\em unstable (i)} and {\em unstable (ii)}). 
The phase 'unstable' (i) and (ii) are respectively characterized by formal solutions $(\rho_{\pm \bolQ}\rightarrow \infty ,\ \rho_{\mp \bolQ}=0)$ and $\rho_{\pm Q}\rightarrow \infty$.
\label{Fig:GL-phase-diag}}
\end{center}
\end{figure}

We summarize the $(\Gamma_1,\Gamma_2))$-phase diagram in 
FIG. \ref{Fig:GL-phase-diag}.   
In section \ref{Sec:J1J2model}, we shall calculate $\Gamma_{1,2}$ for 
a specific model and show that all three possible phases appear.
\section{Low-energy properties}
\label{Sec:low-energy}
\subsection{Effective Lagrangian}
Thus far we have described the general results\cite{Batyev-1}. 
Now we study the low-energy properties of the two phases more closely. 
To this end, it is convenient to 
introduce two independent low-energy modes $a$ and $b$ through 
\begin{equation}
\beta_{l} \sim 
\be^{i{\bf Q}{\cdot}{\bf R}_{l}}a({\bf R})+
\be^{-i{\bf Q}{\cdot}{\bf R}_{l}}b({\bf R}) \; .
\end{equation} 
The mass parameters corresponding to these modes are obtained 
from the low-energy dispersion.  
Defining $\bolk\equiv \bolq{-}\bolQ$, we may expand 
$\epsilon(\bolq)=\epsilon_{\text{min}}+k_i k_j/(2m_{ij})+\cdots $, 
$(i,j=a,b,c)$ 
where the summation over repeated indices is implied.  
The symmetric mass matrix $m_{ij}$ can be diagonalized to give 
a standard form of the dispersion 
$k_i k_j/2m_{ij}=k_i^{\prime 2}/(2m^\prime_{i})\equiv 
\epsilon_{\text{g}}(\bolk^\prime)$. 
We assume $m_{i}>0$ (for all $i$) for the stability of the minima 
at $\mathbf{q}=\pm \mathbf{Q}$.   
Below we omit the prime over $m_j$ for simplicity.  
As a result, we write down the following effective Lagrangian 
with the renormalized interactions:
\begin{equation}
\begin{split}
L_{\text{eff}}\! =&\!\int \! d^3 {\bf x}\!\biggl[ \frac{i}{2} 
(a^\ast\partial _t a\!-\!a\partial _t a^\ast)
\!-\!\frac{|\partial_{j}a|^2}{2m_{j}}
\!+\!\mu a^\ast a \\
&\!+\!\frac{i}{2}(b^\ast\partial _t b-\! b\partial _t b^\ast) 
-\frac{|\partial_{j}b|^2}{2m_{j}}
+ \mu b^\ast b \\
& -\frac{\Gamma_1}{2}(|a|^4\!+\!|b|^4)-\Gamma_2|a|^2|b|^2\biggr] 
\; .
\label{EffL}
\end{split}
\end{equation}
We note that this effective Lagrangian has U(1)${\times}$U(1) symmetry; 
one comes from the axial 
(around the external field) 
symmetry and the other from an {\em emergent} translational symmetry 
which does not exist at the level of the lattice.  
When $\Gamma_1=\Gamma_2$, this U(1)${\times}$U(1)-symmetry gets 
enlarged to U(2).  
On the fully saturated ground state, both $a$ and $b$ bosons have the energy gap $-\mu$.

\subsection{Cone phase}
In the case of $\Gamma_2 > \Gamma_1 >0$, the cone phase appears. 
Only one of the low-energy bosons (say, $a$) condenses and we  
parameterize it as:
$a=\sqrt{\rho+\delta \rho}\, e^{i\theta}$ with 
$\rho_{\bolQ} =\rho=\mu/\Gamma_1$. 
If we integrate out the massive $\delta{\rho}$ fields in the path integral, 
we obtain the following effective Lagrangian:
\begin{equation}
\begin{split}
L_{\rm cone}&=\int d^3{\bf x} 
\left[\left\{ \frac{(\partial_t \theta)^2}{2\Gamma_1}
-\frac{\rho}{2m_{j}}(\partial_{j} \theta)^2\right\}\right. \\
+&\left.\frac{i}{2}(b^\ast\partial _t b-\! b\partial _t b^\ast)
-\frac{|\partial_{j}b|^2}{2m_{j}}
+ \mu(1-\frac{\Gamma_2}{\Gamma_1}) b^\ast b-\frac{\Gamma_1}{2}|b|^4\right],
\end{split}
\end{equation}
From this, we can read off the excitation spectrum of 
the $\theta$-mode as:
\begin{equation}
\Omega_{\text{cone}}(\bolk)=\sqrt{2\mu\epsilon_{\text{g}}(\bolk)}\ ,
\end{equation}
and that of $b$ field acquires a gap $\mu(\frac{\Gamma_2}{\Gamma_1}-1)$.

\subsection{Fan phase}
The low-energy spectrum of the fan phase ($\Gamma_{1}>\Gamma_{2}$, 
$\Gamma_1+\Gamma_2>0$) 
exhibits quite a different behavior, since 
two bosons $a$ and $b$ condense {\em simultaneously} 
and the above-mentioned U(1)${\times}$U(1)-symmetry plays 
a crucial role. 
To see this more clearly, let us parameterize
$a=\sqrt{\rho^{\prime}+\delta \rho_1}\, e^{i\theta_1}$, 
$b=\sqrt{\rho^\prime+\delta\rho_{2}}\, e^{i\theta_2}$ and integrate out 
the massive $\delta{\rho_{1,2}}$ fields in the path integral. 
Then $L_{\text{eff}}$ reduces to:
\begin{equation}
\begin{split}
L_{\rm fan}=&\int d^3{\bf x} 
\left[\left\{ \frac{(\partial_t \theta_u)^2}{2(\Gamma_1+\Gamma_2)}
-\frac{\rho}{2m_{j}}(\partial_{j} \theta_u)^2\right\}\right. \\
&+\left.\left\{ \frac{(\partial_t \theta_v)^2}{2(\Gamma_1-\Gamma_2)}
-\frac{\rho}{2m_{j}}(\partial_{j} \theta_v)^2\right\}\right]\ ,
\end{split}
\label{eqn:EffL-2}
\end{equation}
where we have introduced $\theta_u \equiv(\theta_1+\theta_2)/\sqrt{2}$, 
$\theta_v \equiv (\theta_1-\theta_2)/\sqrt{2}$.  
Now the meanings of the two angular variables appearing in 
eq.(\ref{superfan}) are clear; the field $\theta_u$ corresponds to 
the Goldstone mode associated with the spontaneous breaking of  
rotational symmetry in the $x$-$y$ plane, while the other $\theta_v$ 
describes the translational motion of the fan along $\bolQ$ 
({\em phason} of the spin ($S^{z}$) density wave). 
The excitation spectrum of the $\theta_u$-mode is readily obtained 
from (\ref{eqn:EffL-2}) as
\begin{equation}
\Omega_u({\bf k}) = \sqrt{2(\Gamma_1 + \Gamma_2)
\rho^\prime\epsilon_{\text{g}}(\bolk)}
= \sqrt{2\mu\,\epsilon_{\text{g}}(\bolk)} \; 
(=\Omega_{\text{cone}}(\bolk))\ .\label{GSmode}
\end{equation}
This gapless excitation does not exist in the fan phase appearing 
in classical models with easy plane anisotropy\cite{Nagamiya}. 
Similarly, the excitation spectrum related to $\theta_v$ is given by
\begin{equation}
\Omega_v({\bf k}) \!=\! \sqrt{2(\Gamma_1-\Gamma_2)\rho^\prime
\epsilon_\text{g}({\bf k})}=\sqrt{2
\frac{\Gamma_1-\Gamma_2}{\Gamma_1+\Gamma_2}
\mu\epsilon_\text{g}({\bf k})}.\label{phason}
\end{equation}
At the transition point $\Gamma_{1}=\Gamma_{2}$ from the cone to 
the fan, the phonon velocity of the $\theta_{v}$-mode vanishes 
indicating an instability in the translational mode. 

\section{Effects of commensurability}
\label{Sec:Commensurability}
In this section, we consider effects of commensurability 
on the ground state.   Since our system is defined 
on a lattice, any types of interactions which are allowed by the
symmetry may be added to the effective Lagrangian (\ref{EffL}). 
Specifically, we require invariance under 
\begin{equation}
\begin{split}
& \text{(i) global U(1):} \quad (a,b)\mapsto \be^{i\theta}(a,b) \\ 
& \text{(ii) lattice translation:} \quad  
a\mapsto a\,\be^{i\bolQ{\cdot}\boldsymbol{\delta}} \; , \;\;
b\mapsto b\,\be^{-i\bolQ{\cdot}\boldsymbol{\delta}} \\ 
& (\boldsymbol{\delta}\text{: lattice period}) \; .
\end{split}
\end{equation}  
For generic incommensurate values 
of $\bolQ$, the effective Lagrangian (\ref{eqn:EffL-2})  
correctly describes the low-energy physics.   
If $Q$ is rational (i.e. $Q=Q_{0} = \pi n/l$: $l$,$n$ are coprime), 
on the other hand, the following terms in general appear 
to break the translational U(1) symmetry explicitly:
\begin{equation}
\begin{split}
\frac{1}{2}\Gamma_3
(a^{\dagger l}b^l + b^{\dagger l}a^{l})
& \simeq \Gamma_3\rho^{\prime l}\cos l(\theta_1-\theta_2) \\
& = \Gamma_{3}\rho^{\prime} \cos(\sqrt{2}\,l\,\theta_{v}) \; .
\end{split}
\label{gamma3}
\end{equation}
If we treat the problem in a classical manner, we can imagine 
an infinite sequence of crystalline phases with superfluid order 
(spin analogue of supersolids).  

However, if we take into account quantum fluctuation, 
this devil's staircase structure is destroyed 
and only a finite number of commensurate phases survive\cite{PerBak}. 
To see this explicitly, we redefine the boson operator as:
\begin{equation}
\begin{split}
& a(\mathbf{R})\mapsto
\be^{-i\delta\bolQ{\cdot}\mathbf{R}}a(\mathbf{R}) \; , \;\;
b(\mathbf{R})\mapsto \be^{+i\delta\bolQ{\cdot}\mathbf{R}}b(\mathbf{R})
\\
& (\mathbf{\delta Q}\equiv (0,0,Q- Q_{0})) \; .
\end{split}
\end{equation} 
Then, the effective Lagrangian is given as, 
\begin{equation}
\begin{split}
&L_{\text{eff}2} =\int d^3 {\bf x}\left[ \frac{i}{2} 
(a^\ast\partial _t a-a\partial _t a^\ast)
-\frac{|(-i\partial_{j}-\delta Q_{j})a|^2}{2m_{j}}\right. \\
&+ \mu a^\ast a 
+\frac{i}{2}(b^\ast\partial _t b-b\partial _t b^\ast)
-\frac{|(-i\partial_{j}+\delta Q_{j})b|^2}{2m_{j}}
+ \mu b^\ast b\\
&-\frac{\Gamma_1}{2}(|a|^4 + |b|^4)
-\Gamma_2|a|^2|b|^2\left. -\frac{\Gamma_3}{2}(a^{\ast l}
b^l + b^{\ast l}a^l) \right].
\end{split}
\end{equation}

From now, we concentrate on the fan phase where both 
$a$ and $b$ condense.  
As before, we integrate out the massive $\delta{\rho_{1,2}}$ fields 
in the path integral, and ignore the terms which do not matter 
for low-energy physics when the superfluid density $\rho$ is 
dilute and $l\geq 3$. 
For $l=2$ case, $\Gamma_3$ term is the same order 
as $\Gamma_1$ and $\Gamma_2$ term in $E/N$ and therefore fan phase 
appears for\cite{Jackeli-Zhito} $\Gamma_1+|\Gamma_3|>\Gamma_2$.
Thus, our approximation may not be justified for $l=2$ case. 
Now, $L_{\text{eff2}}$ reads, 
\begin{equation}
\begin{split}
&L_\text{eff2}^\prime\approx \int d^3{\bf x}\Bigl[
\left\{ \frac{(\partial_t \theta_u)^2}{2(\Gamma_1+\Gamma_2)}
-\frac{\rho}{2m_{j}}(\partial_{j} \theta_u)^2\right\}\\
+&\! \left\{ \!\frac{(\partial_t
 \theta_v)^2}{2(\Gamma_1-\Gamma_2)}\!-\!\frac{\rho}{2m_{j}}
( \partial_{j} \theta_v\!-\!\sqrt{2}\delta Q_{j})^2\!
-\!\Gamma_3 \rho^l\cos \sqrt{2}l \theta_v \! \right\}\Bigr].
\end{split}
\end{equation}
Classically, if $\delta Q$ is small,  
the $\Gamma_3$-term seems to pin the translation mode $\theta_v$ 
at the expense of the elastic energy. 
If $\theta_v$ is pinned, on the other hand, 
the gapped zero-point fluctuations 
around the pinned value yield the (positive) quantum correction 
to the ground state energy and thereby a soliton lattice 
with gapless excitations\cite{soliton} may be favored.   
From eq.(14) in Ref.\onlinecite{PerBak}, 
an incommensurate soliton lattice is stable for any $\delta Q$ 
if the following inequality is satisfied: 
\begin{equation}
l^2> \text{Min}\left[
16\left( \frac{\rho}{m_i(\Gamma_1-\Gamma_2)}\right)^{1/2}
\right] \ ,\label{inequality}
\end{equation}
where ${\rm Min}[\cdots ]$ means that the minimum value with 
respect to $i=a,b,c$ should be taken.  
Therefore, at least in the dilute gas limit, i.e., just below the saturation 
field, a commensurability locking does not occur for $l\geq 3$. 
When $\rho$ grows further, eq.(\ref{inequality}) may be violated 
for some small commensurability $l$ and the locking occurs; 
the pitch $Q$ is locked to its commensurate value $Q_{0}$ 
until $\delta Q$ exceeds the critical value\cite{PerBak} 
\begin{equation}
\delta Q_{\text{c}}^2=(8/\pi^2)2m_c|\Gamma_3|
\rho^{l-1}\{
1-(l^2/16)\sqrt{m_c(\Gamma_1-\Gamma_2)/\rho}\}^2.
\end{equation}
It is interesting to see that the first term, which is obtained 
by classical calculation, has the same character 
as a classical fan phase in an easy plane, which has a width 
proportional to\cite{Cadorin-Yokoi} 
$\delta Q^2\propto |\text{H}-\text{H}_{\text{c}}|^{l-1}$.

Before concluding this section, we would like to give 
a remark on the validity of our treatment.  
Above discussions assume the dilute-gas limit, 
where the scattering length is much smaller than the average 
interatomic distance $\rho^{-1/3}$. Specifically, 
our approximation is valid when 
$\Gamma_i (m_am_bm_c\rho)^{1/3} \ll 1$ is satisfied for $i=1$ or $2$. 
\section{Coupled $J_1$-$J_2$ model}
\label{Sec:J1J2model}
\subsection{Phases of a single $J_1$-$J_2$ chain}
Before presenting our results for a 3D model ($J_1$-$J_2$-$J_3$ model), 
let us briefly review the known results for the $S=1/2$ $J_1$-$J_2$ 
chain (the case with $J_3=0$) and discuss the connection 
to the phases found in Sec.\ref{Sec:General}.  
In the case $J_{1}>0$, near saturation, two dominant phases are 
found\cite{1D-J1-J2-AF}:
(i) `chiral phase (VC)' with finite vector chirality parallel 
to the magnetic field\cite{Kolezhuk-V,1D-J1-J2-AF} 
and (ii) `TL2' phase where the system is 
described by two Tomonaga-Luttinger (TL) 
liquids\cite{Okunishi-T,1D-J1-J2-AF}. 
Obviously, the former turns, after switching on an interchain coupling, into 
the cone phase. A close inspection of the two gapless TL modes 
near saturation tells us that the TL2 phase should evolve into 
the fan phase in three dimensions where we have two Goldstone 
modes.  Yet another dominant phase `TL1', for which a single-component 
TL gives a good description\cite{Kolezhuk-V,1D-J1-J2-AF}, 
is located in a region where 
we expect a more conventional single-component BEC at 
$\mathbf{Q}=(0,0,\pi)$ ($J_3<0$) or $(\pi,\pi,\pi)$ ($J_3>0$).   

The ferromagnetic side $J_{1}<0$ is much more subtle as we expect  
BECs of $n$-bound magnon states ($n\geq 2$) to occur.  
In one-dimension ($J_3=0$), on top of the VC phase described above, 
various phases related to bound $n$-magnons ($2\leq n \leq 4$) 
have been found\cite{1D-J1-J2-FM}; 
(a) TL phases of 2-magnon bound states `nematic' and `SDW$_2$', 
whose dominant correlation occur in respectively superfluid- 
and SDW channel, (b) 3-magnon TL `triatic' and `SDW$_3$' 
(the meanings of them are evident) and (c) `quartic' corresponding 
to 4-magnon bound states.  
Our dilute-gas analysis predicts that inside the domes of the attraction-dominant phase (phase (iii) in Fig.\ref{Fig:quasi1D}) 
one-magnon BEC becomes unstable toward various kinds of 
magnon bindings as has been discussed 
in Sec.\ref{attraction-dominant} and we may expect that 
the above $n$-magnon-based phases correspond to the attraction-dominant 
phase. 
We also study the stability of the 2-magnon bound state by the traditional approach\cite{Mattis} and discuss in later part of this section.

\subsection{3D phase diagram}
Having established the formalism, 
we now consider a frustrated spin-1/2 model on a simple cubic lattice 
whose Hamiltonian is given by
\begin{equation}
H\!=\!\!\!\!
\sum_{\bolr,i=a,b}\!\!\!\! \left\{J_1 {\bf S}_{\bolr}{\cdot}
{\bf S}_{\bolr+\hat{\bole}_{c}}
\!+\!J_{2}{\bf S}_{\bolr}{\cdot}{\bf S}_{\bolr+2\hat{\bole}_{c}}
\!+\!J_3{\bf S}_{\bolr}{\cdot}{\bf S}_{\bolr+\hat{\bole}_{i}}\right\} ,
\end{equation}
where we label the three crystal axes by $(a,b,c)$ 
and the spiral vector $\bolQ$ is pointing the $c$-direction.   
The $J_{1}$-$J_{2}$ chains are running in the $c$-direction 
and $J_{3}$ controls the coupling among adjacent chains.  

If we replace the spin-1/2s by hardcore bosons, 
we obtain the bosonic Hamiltonian (\ref{Hboson}) with $\epsilon(q)$ given by:
\begin{equation}
\epsilon(q)=J_1 \cos q_c+J_2\cos 2q_c + J_3(\cos q_a+\cos q_b)\ .
\end{equation}

\begin{figure}[ht]
\begin{center}
\includegraphics[scale=0.55]{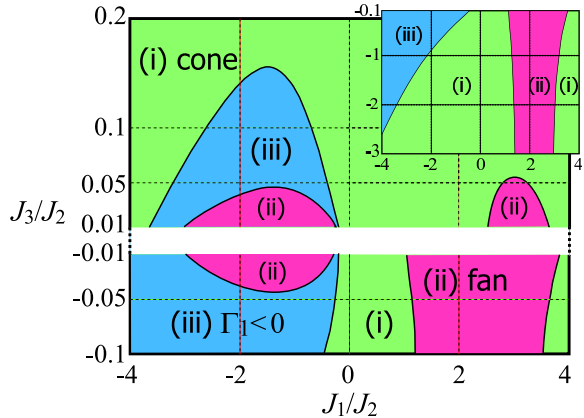}
\caption{(Color online) 
Phase diagram slightly below saturation 
($\text{H}\lesssim \text{H}_{\text{c}}$) mapped out 
in $(J_{1},J_{3})$-plane 
obtained from the one-magnon-BEC approach 
($J_{2}(>0)$ is used to set the energy unit). 
Note that only the region 
$-4 \le J_{1}/J_{2} \le 4$, where cone structure with 
incommensurate ${\bf Q}$ is expected classically, is shown.   
Two Bose-condensed phases (i) {\em cone phase} and 
(ii) {\em coplanar fan phase} as well as 
the phase-(iii) which is characterized by $\Gamma_{1}<0$ or 
$\Gamma_{1}+\Gamma_{2}<0$ are shown.  
The region $|J_{3}|/J_{1}\ll 1$ is omitted (see the text). 
Inset: The same phase diagram for the large negative 
interchain coupling ($J_{3}<-0.1$). %
\label{Fig:quasi1D}}
\end{center}
\end{figure}

The mass parameters are given by $m_a=m_b=1/|J_3|$, and
$m_c=1/(4J_2-{J_1^2}/{4J_2})$. 
The wave number ${\bf Q}$ characterizing the condensate is given 
either by $\bolQ=(0,0,Q)$ ($J_{3}<0$) or 
by $\bolQ=(\pi,\pi,Q)$ ($J_{3}>0$) where 
$Q=\arccos (-{J_1}/{4J_2})$.  
We solved eq.(\ref{laddereq}) to determine the spin structure 
of our $J_{1}$-$J_{2}$-$J_{3}$ model 
(see Appendix~\ref{Ap-1} for the details). 
As a result, we obtained the phase diagram  
shown in FIG.\ref{Fig:quasi1D}.

On top of the ordinary cone phase, strong quantum fluctuation 
in $S=1/2$ systems stabilizes two new phases: 
the coplanar fan ((ii)) and the attraction-dominant phase ((iii)). 
In the phase-(ii), both gauge symmetry and translation symmetry 
are broken simultaneously and as a consequence we have two 
different low-energy (Goldstone) modes 
$\Omega_{u}(\bolk)$ (eq.(\ref{GSmode})) 
and $\Omega_{v}(\bolk)$ (eq.(\ref{phason})). 
In the phase-(iii), strong attraction may imply instabilities 
toward other phases e.g. conventional ferromagnetic one or 
more exotic multipolar ones\cite{1D-J1-J2-FM}.  
For $J_3 \rightarrow 0$, low-energy quantum fluctuation 
destabilizes $\Gamma$  
and our approach cannot be extended to $J_3=0$ 
continuously ($\Gamma$ becomes $O(J_3^{1/2}$) and 
$\Gamma_1 \rightarrow \Gamma_2$ at the leading order in $J_3$).

For the ferromagnetic $J_{3}$, the cone phase (region-(i)) 
gets wider and wider and the boundary 
between the cone- and the phase-separated phases approaches 
the classical phase boundary $J_{1}/J_{2}=-4$ as $|J_{3}|$ is increased 
(see the inset of FIG.\ref{Fig:quasi1D}).   
This is easily understood since for very large negative 
$J_{3}$ all spins sitting on each $ab$-plane behave like 
{\em a single large spin} to which classical analysis is applicable 
and the system may be thought of as a {\em single} chain running in 
the $c$-direction. 
For the antiferromagnetic coupling ($J_{3}>0$), these novel phases ((ii)
and (iii)) appear only in the weak-coupling ($J_{3} \ll J_{1,2}$) 
region. 

\begin{figure}[ht]
\begin{center}
\includegraphics[scale=1.0]{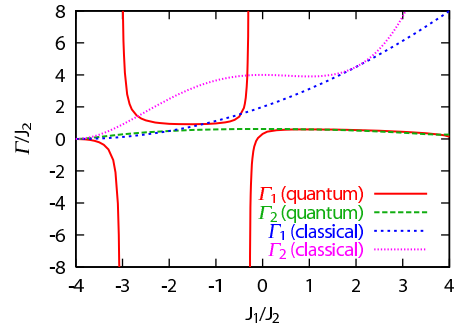}
\caption{(Color online) 
$\Gamma_1$ and $\Gamma_2$ for $J_3/J_2=-0.01$ plotted as a function 
of $J_{1}/J_{2}$. 
Also plotted are $\Gamma_{i}$ of classical 1D-chain ($J_{3}=0$), 
where we use the spin-wave expansion around the saturated phase 
and retain only the leading term of the 4-point interaction 
in $S$ as in Ref.\onlinecite{Nikuni-Shiba-2}. 
In the classical case, the cone phase is always stable except at $J_1/J_2=2$.
In the quantum case ($J_3/J_2=-0.01$), $\Gamma_1$ has two poles at
 the boundaries between the phase (ii) and (iii) in FIG.\ref{Fig:quasi1D}.
\label{Fig:gamma}}
\end{center}
\end{figure}

To see the possible magnon binding more clearly, we plot $\Gamma_i$ in Fig.\ref{Fig:gamma}.  
Although $\Gamma$s behave regularly in the most part of the phase diagram, 
$\Gamma_1$ has poles on the boundary between the fan phase (ii) and 
the phase-(iii) as is seen in Fig.\ref{Fig:gamma}. 
This implies that near the boundary between the phase-(ii) and (iii) 
the interaction among bosons becomes singularly large 
which may lead to new phases. 
Actually, a pole of a interaction between two particles in general  
imply an existence of stable bound states. 
Thus, the one-magnon-BEC approach is not sufficient to see the ground state
 near this boundary.

\begin{figure}[ht]
\begin{center}
\includegraphics[scale=0.55]{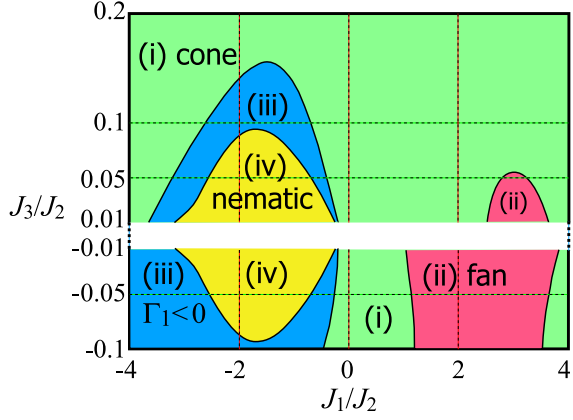}
\caption{(Color online) 
The same phase diagram as in Fig.\ref{Fig:quasi1D} when the 2-magnon bound state is taken into account. 
The phase (iv) is characterized by the condensation of the 2-magnon bound state and leads to the nematic order in the transverse direction. 
\label{Fig:quasi1Dnem}}
\end{center}
\end{figure}

To highlight this point, 
we study the instability of the 2-magnon bound state. 
On the fully saturated ground state $\ket{FM}$, the wave function of the 2-magnon bound state is given by $\sum_{i,j} \psi(i,j)S^+_i S^+_j\ket{FM}$, and the energy of this wave function can be exactly obtained by solving the two-body Schr\"{o}dinger equation\cite{Mattis}. 
If the gap of the bound state closes earlier than that of the one magnon, the bound-magnon BEC will occur, and the nematic order emerges in the transverse direction.
We show the region of the stable bound state in Fig.\ref{Fig:quasi1Dnem}. As a result, the nematic phase completely masks the fan phase which would have appeared on the ferromagnetic side $J_{1}<0$. 
At a rough estimate, for $-2.7\lesssim J_1/J_2$, the lowest-mode of the bound state is commensurate and for $J_1/J_2\lesssim -2.7$ the one is incommensurate in the same way as in the 1D $J_1$-$J_2$ chain\cite{Chubukov,Comm}.

Detailed results on the bound magnons will be reported elsewhere\cite{UTM}.

\section{Summary}
By using the dilute-Bose-gas technique, 
we studied the high-field magnetic structures and low-energy excitations of 
three-dimensional quantum ($S=1/2$) helimagnet. 
The method used is asymptotically exact when magnetization is close to 
saturation (we gave a criterion of `proximity' in the end of section IV) and 
enables us to obtain reliable results for {\em three-dimensional} frustrated 
systems. 

In Sec.~\ref{Sec:General}, we discussed various phases emerging from the BEC of hard-core magnons slightly below the saturation field. 
Although only the cone phase is expected 
in the classical helimagnets\cite{Nagamiya}, 
quantum fluctuation can stabilize the fan or attraction dominant phases 
as well 
according to the renormalized interactions $(\Gamma_1,\Gamma_2)$. 

Then, the low-energy excitations of the cone and the fan phases were 
studied in Sec.~\ref{Sec:low-energy}. 
The hallmark of quantum helimagnets is that one has two low-energy modes 
at ${\bf k}=\pm{\bf Q}$ and the low-energy physics is described 
by the effective Lagrangian with U(1)$\times$U(1) symmetry;
one comes from the axial (around the external field) 
symmetry and the other from an {\em emergent} translational symmetry. 
In the cone phase, only one of the two bosons condenses and there 
is one gapless Goldstone mode.
Meanwhile, the fan phase breaks both symmetries and has two types of 
gapless Goldstone modes.

In Sec.~\ref{Sec:Commensurability}, we considered effects of 
commensurability on the helical modulation vector $Q$. 
If $Q$ is rational (i.e. $Q=Q_{0} = \pi n/l$: $l$,$n$ are coprime), 
additional interactions 
appear in the effective Lagrangian and favors the gapped commensurate phase. 
By examining the correlation due to the gapped zeropoint fluctuations, 
we found, for $l \geq 3$, that slightly below the saturation field quantum fluctuation destroys 
the commensurate order. 

We studied a concrete three-dimensional $S=1/2$ model 
($J_1$-$J_2$-$J_3$ model) 
in Sec.~\ref{Sec:J1J2model} 
and mapped out its (high-field) phase diagram in Figs.\ref{Fig:quasi1D} and 
\ref{Fig:quasi1Dnem}. 
An analysis assuming the single-magnon BEC predicts the existence of 
a fairly large region where single-magnon BECs may be unstable (phase-(iii)) 
as well as the cone and the fan shown in Fig.\ref{fig:cone-fan}. 
To get an insight into the nature of this `phase', 
we consider the possibility of a BEC of {\em two-magnon} bound states, 
which leads to the nematic order in the transverse direction. 
In fact, in a large portion of the phase-(iii) (a region marked as (iv) 
in Fig.\ref{Fig:quasi1Dnem}), we have a stable two-magnon 
bound states which condense first when the external field is 
decreased.  In the 1D $J_1$-$J_2$ model (i.e. $J_3=0$), it is known that 
one has multiple-magnon bound states up to four-body\cite{1D-J1-J2-FM} 
and some parts of the new phase ({\em  nematic} (iv) 
in Fig.\ref{Fig:quasi1Dnem}) 
could be replaced by the condensed phases of these bound states. 

Finally, we comment on the relevance of our study to real systems. 
Interests in multiferroicity sparked an intensive study of various 
helimagnetic materials, among which one can find many examples of 
coupled $J_1$-$J_2$ chains. For example, LiCuVO$_4$ is characterized 
by edge-sharing CuO$_2$ plaquettes and may be modeled  
by the $S=1/2$ $J_1$-$J_2$ chain with negative $J_1$. 
Neutron diffraction and {\em ab initio} calculations 
suggested\cite{LiCuVo4}, as well as $J_1$ and $J_2$, 
various kinds of interchain interactions $J_{3},..J_{6}$.  
Although the stacking of $J_1$-$J_2$ chains is different 
from what is assumed here, our method can be readily generalized 
to include more realistic cases and we hope our approach will 
shed some light on magnetism of these quantum helimagnets.

\begin{acknowledgements}
We thank S.~Furukawa, A.~Furusaki, T.~Hikihara, T.~Momoi, 
T.~Nishino, M.~Sato, and N.~Shannon for discussions. 
One of us (K.T.) was supported by Grant-in-Aid 
for Scientific Research (C) 20540375 and that on 
Priority Areas ``Novel States of Matter Induced by Frustration'' 
(No.19052003) from MEXT, Japan.  
This work was also supported by the Grant-in-Aid for the Global 
COE Program 
``The Next Generation of Physics, Spun from Universality 
and Emergence" from MEXT of Japan.
\end{acknowledgements}

\appendix
\section{how to treat the ladder diagram}
\label{Ap-1}
For convenience, we briefly summarize the method of calculating $\Gamma$.
To obtain $\Gamma_{1}=\Gamma_{\bolq=0}(\bolk_1=\bolQ,\bolk_2=\bolQ)$ and $\Gamma_2=\Gamma_0(\bolQ,-\bolQ)+\Gamma_{-2\bolQ}(\bolQ,-\bolQ)$, we solve the following integral equation in the case of $\bolk_{1,2}=\pm \bolQ\ \ (\omega(\bolk_{1,2})=0)$:
\begin{equation}
\Gamma_{\bolq}(\bolk_1,\bolk_2)=V_\bolq-\frac{1}{N}\sum_{\bolq^\prime} \frac{\Gamma_{\bolq^\prime}(\bolk_1,\bolk_2)V_{\bolq-\bolq^\prime}}{\omega(\bolk_1+\bolq^\prime)+\omega(\bolk_2-\bolq^\prime)}\ ,\label{GammaEqA}
\end{equation}
where $V_{\bolq} =2(\epsilon(\bolq)+U)$. In what follows, we do not write the argument $\bolk_{1,2}$ of $\Gamma$ explicitly, 
and denote $\frac{1}{N}\sum_q$ as $\VEV{\ }$. Since $\VEV{\epsilon}=0$, we sum up the both side of (\ref{GammaEqA}) with respect to $\bolq$ and obtain,
\begin{equation}
\VEV{\Gamma}=2U\left(1-\frac{1}{N}\sum_{\bolq^\prime}\frac{\Gamma_{\bolq^\prime}}{\omega(\bolk_1+\bolq^\prime)+\omega(\bolk_2-\bolq^\prime)}\right),\label{UGamma}\\
\end{equation}
Using this equation, (\ref{GammaEqA}) is simplified to,
\begin{equation}
\begin{split}
&\Gamma_{\bolq}=2\epsilon(\bolq)+\VEV{\Gamma}\\
&\hspace{0.8cm}-\frac{1}{N}\sum_{\bolq^\prime} \frac{2\epsilon(\bolq-\bolq^\prime)}{\omega(\bolk_1+\bolq^\prime)+\omega(\bolk_2-\bolq^\prime)}\Gamma_{\bolq-\bolq^\prime}\ .\label{GammaUnasi}
\end{split}
\end{equation}
Additionally, if we assume the limit $U\rightarrow \infty$, eq.(\ref{UGamma}) reads:
\begin{equation}
1-\frac{1}{N}\sum_\bolq\frac{\Gamma_{\bolq^\prime}}{\omega(\bolk_1+\bolq^\prime)+\omega(\bolk_2-\bolq^\prime)}=0\label{gammaeq1} \ .
\end{equation}
Now, the problem is reduced to solve (\ref{GammaUnasi}) and (\ref{gammaeq1}) simultaneously, which are free from the infinite term $U$. Next, we expand $\Gamma_\bolq$ in lattice harmonics. 
Since
\begin{equation}
\begin{split}
\epsilon(\bolq-\bolq^\prime)=&J_1(\cos q_c \cos q_c^\prime +\sin q_c\sin q_c^\prime)\\
+&J_2(\cos 2q_c \cos 2q_c^\prime +\sin 2q_c\sin 2q_c^\prime)\\
+&J_3(\cos q_a \cos q_a^\prime +\sin q_a\sin q_a^\prime\\
&+\cos q_b \cos q_b^\prime +\sin q_b\sin q_b^\prime)\ ,
\end{split}
\end{equation}
we introduce
\begin{equation}
\begin{split}
\Gamma_\bolq&=\VEV{\Gamma}+J_1A_1\cos q_c+J_1A_2\sin q_c +J_2A_3\cos 2q_c\\
&+J_2A_4\sin 2q_c+J_3A_5(\cos q_a+\cos q_b)\ .
\end{split}
\end{equation}
We note that $\VEV{\Gamma}$ and $A_i$ are independent of $\bolq$, but depend on $(\bolk_1,\bolk_2)$ implicitly. 
If we substitute this, eq.(\ref{GammaUnasi}) reduces to, 
\begin{widetext}
\begin{equation}
\begin{split}
&\left(J_1A_1+\frac{1}{N}\sum_{\bolq^\prime} \frac{2J_1\cos q_c^\prime}{\omega(\bolk_1+\bolq^\prime)+\omega(\bolk_2-\bolq^\prime)}\Gamma_{\bolq^\prime}-2J_1\right)\cos q_c
 +\left(J_1A_2+\frac{1}{N}\sum_{\bolq^\prime} \frac{2J_1\sin q_c^\prime}{\omega(\bolk_1+\bolq^\prime)+\omega(\bolk_2-\bolq^\prime)}\Gamma_{\bolq^\prime}\right)\sin q_c\\
+&\left(J_2A_3+\frac{1}{N}\sum_{\bolq^\prime} \frac{2J_2\cos 2q_c^\prime}{\omega(\bolk_1+\bolq^\prime)+\omega(\bolk_2-\bolq^\prime)}\Gamma_{\bolq^\prime}-2J_2\right)\cos 2q_c
+\left(J_2A_4+\frac{1}{N}\sum_{\bolq^\prime} \frac{2J_2\sin 2q_c^\prime}{\omega(\bolk_1+\bolq^\prime)+\omega(\bolk_2-\bolq^\prime)}\Gamma_{\bolq^\prime}\right)\sin 2q_c\\
+& \left(J_3A_5+\frac{1}{N}\sum_{\bolq^\prime} \frac{J_3(\cos q_{a}^\prime+\cos q_b^\prime)}{\omega(\bolk_1+\bolq^\prime)+\omega(\bolk_2-\bolq^\prime)}\Gamma_{\bolq^\prime}-2J_3\right)(\cos q_a+\cos q_b) =0\ ,\label{gammaeq26}
\end{split}
\end{equation}
\end{widetext}
where we use the relation
\begin{equation}
\frac{1}{N}\sum_{\bolq^\prime} \frac{\sin q_{x,y}^\prime}{\omega(\bolk_1+\bolq^\prime)+\omega(\bolk_2-\bolq^\prime)}\Gamma_{\bolq^\prime}=0\ .
\end{equation}
To satisfy the eq.(\ref{gammaeq26}) for arbitrary $\bolq$, the coefficients of trigonometric function of $\bolq$ must be $0$. 
For convenience, we define
\begin{equation}
\tau_{ij}(k_1,k_2)=\frac{1}{N}\sum_{q^\prime} \frac{T_i(q^\prime)T_j(q^\prime)}{\omega(k_1+q^\prime)+\omega(k_2-q^\prime)}\ ,
\end{equation}
where 
\begin{equation}
{\bf T}(q)=(1,\ \cos q_c,\ \sin q_c,\ \cos 2q_c,\ \sin 2q_c,\ \cos q_a+\cos q_b)\ .
\end{equation}
Then, (\ref{gammaeq1}) and (\ref{gammaeq26}) are put together into
\begin{widetext}
\begin{equation}
\left(
\begin{array}{cccccc}
\tau_{11} & J_1\tau_{12} & J_1\tau_{13} & J_2\tau_{14} & J_2\tau_{15} & J_3\tau_{16}\\
2\tau_{21} & 1+2J_1\tau_{22} & 2J_1\tau_{23} & 2J_2\tau_{24} & 2J_2\tau_{25} & 2J_3\tau_{26}\\
2\tau_{31} & 2J_1\tau_{32} & 1+2J_1\tau_{33} & 2J_2\tau_{34} & 2J_2\tau_{35} & 2J_3\tau_{36}\\
2\tau_{41} & 2J_1\tau_{42} & 2J_1\tau_{43} & 1+2J_2\tau_{44} & 2J_2\tau_{45} & 2J_3\tau_{46}\\
2\tau_{51} & 2J_1\tau_{52} & 2J_1\tau_{53} & 2J_2\tau_{54} & 1+2J_2\tau_{55} & 2J_3\tau_{56}\\
\tau_{61} & J_1\tau_{62} & J_1\tau_{63} & J_2\tau_{64} & J_2\tau_{65} & 1+J_1\tau_{66}\\
\end{array}
\right)
\left(
\begin{array}{c}
\VEV{\Gamma}\\
A_1\\
A_2\\
A_3\\
A_4\\
A_5\\
\end{array}
\right)
=
\left(
\begin{array}{c}
1\\
2\\
0\\
2\\
0\\
2\\
\end{array}
\right)
\end{equation}
\end{widetext}
This equation can be solved by calculating $\tau_{ij}$ numerically. 
If we evaluate at $(\bolk_1,\bolk_2)=(\bolQ,\bolQ)$, $A_2=A_4=0$ due to the symmetry and we obtain,
\begin{equation}
\Gamma_1=\VEV{\Gamma}+J_1A_1+J_2A_3+2J_3A_5\ .
\end{equation} 
If we evaluate at $(\bolk_1,\bolk_2)=(\bolQ,-\bolQ)$, we obtain
\begin{equation}
\begin{split}
\Gamma_2 & =\left[\VEV{\Gamma}+J_1A_1+J_2A_3+2J_3A_5\right]\\
& +[\VEV{\Gamma}+J_1A_1\cos (-2Q)+J_1A_2\sin (-2Q)\\
&+J_2A_3\cos (-4Q)+J_2A_4\sin (-4Q)+2J_3A_5]\ ,
\end{split}
\end{equation}
Although above we review the straightforward method, 
we can calculate $\Gamma_2$ more simply if we introduce,
\begin{equation}
\begin{split}
\Gamma_{\bolq-\bolQ}(\bolQ ,-\bolQ)&+\Gamma_{-\bolq-\bolQ}(\bolQ ,-\bolQ)\\
=&2\VEV{\Gamma}+J_1A_1^\prime\cos q_c+J_2A_2^\prime\cos 2q_c\\
&+J_3A_3^\prime(\cos q_a+\cos q_b)\ .
\end{split}
\end{equation}
The following procedure is the same as in the former case, and $\Gamma_2$ is given by,
\begin{equation}
\Gamma_2=2\VEV{\Gamma}+J_1A_1^\prime\cos Q+J_2A_2^\prime\cos 2Q-2|J_3|A_3^\prime\ .
\end{equation}



\begin{thebibliography}{101}
\bibitem{footnote1}
We will discuss the validity of our dilute-Bose-gas approach 
in the end of section \ref{Sec:Commensurability}.
\end{thebibliography}
\end{document}